\documentclass[aps,prl,preprint,superscriptaddress]{revtex4-1}

\usepackage{graphicx}

\begin{document}


\title{Ultra strong coupling regime and plasmon-polaritons in parabolic semiconductor quantum wells}


\author{Markus Geiser}
\email[]{mgeiser@ethz.ch}
\affiliation{Institute for Quantum Electronics, ETH Zurich, Wolfgang-
Pauli-Strasse 16, 8093 Zurich, Switzerland.}
\author{Fabrizio Castellano}
\affiliation{Institute for Quantum Electronics, ETH Zurich, Wolfgang-
Pauli-Strasse 16, 8093 Zurich, Switzerland.}
\author{Giacomo Scalari}
\affiliation{Institute for Quantum Electronics, ETH Zurich, Wolfgang-
Pauli-Strasse 16, 8093 Zurich, Switzerland.}
\author{Mattias Beck}
\affiliation{Institute for Quantum Electronics, ETH Zurich, Wolfgang-
Pauli-Strasse 16, 8093 Zurich, Switzerland.}
\author{Laurent Nevou}
\affiliation{Institute for Quantum Electronics, ETH Zurich, Wolfgang-
Pauli-Strasse 16, 8093 Zurich, Switzerland.}
\author{J\'{e}r\^{o}me Faist}
\email[]{jerome.faist@phys.ethz.ch}
\affiliation{Institute for Quantum Electronics, ETH Zurich, Wolfgang-
Pauli-Strasse 16, 8093 Zurich, Switzerland.}


\date{\today}

\begin{abstract}
Ultra strong coupling is studied in a modulation-doped parabolic potential well coupled to an inductance-capacitance resonant circuit. In this system, in accordance to Kohn's theorem, strong reduction of the energy level separation caused by the electron-electron interaction compensates the depolarization shift. As a result, a very large ratio of 27\% of the Rabi frequency to the center resonance frequency as well as a polariton gap of width $2 \pi \times 670$GHz are observed, suggesting parabolic quantum wells as the system of choice in order to explore the ultra-strong coupling regime. \end{abstract}


\maketitle

Recently, there is growing interest in the study of light-matter coupling in the so-called strong coupling regime \cite{weisbuch, wallraff, bloch, imamoglu, ladder}, where the absorption length is so large that energy can be exchanged between the light and matter excitations many times before the system decays back into its ground state \cite{cohen}. Such a situation can be reached in optical microcavities where the frequency of the empty cavity matches the one of an electronic transition. The exchange between the two excitations occurs at a rate equal to twice the Rabi frequency $\Omega_R$, the value of which is a fundamental measure of the coupling strength of the two systems.
As a result, light and matter cannot be considered as separate entities any more, and the energy spectrum of the coupled system is split into two well separated resonances, the upper and lower polaritons,  which are collective excitations of the coupled light-matter system  having an energy separation of $2\Omega_R$.

Special interest was devoted to electronic systems where $\Omega_R$ is a substantial fraction of the energy of the uncoupled electronic excitation. A range of new physical phenomena has been predicted in this so-called ultra strong coupling regime, including the parametric generation of non-classical light upon strong modulation of carrier density\cite{ciuti05, ultrastrong1, ultrastrong2}. THz intersubband transitions in quantum wells are extremely attractive systems for the observation of strong ~\cite{dini, geiser, todorov} and ultra strong~\cite{ huber, ultrastrong3, todorov2010} light-matter coupling, as an electron gas with high density can be created in the ground state by doping and as $\Omega_R$ scales naturally with the square root of the electron density. In fact, electron-electron interactions are naturally expected to play an important role in such devices with a large electron density.  

In particular, Todorov et al.~\cite{todorov2010} have reported a metallic microcavity containing a multiquantum well system that achieved a ratio $\Omega_R/\tilde{\omega} = 0.2$, where $\tilde{\omega}$ is the frequency of the intersubband plasmon without cavity. The large value of the interaction $\Omega_R$, achieved using heavily doped quantum wells, was accompanied by a blueshift of the frequency of the bare intersubband transition $\omega_{12}$ to $\tilde{\omega}$ caused by depolarization shift \cite{ando} as shown in a model in the dipolar gauge instead of the Coulomb gauge. Because of the similar role of the plasma frequency in the depolarization shift and the Rabi frequency $\Omega_R$, a large depolarization shift was identified as a feature automatically accompanying a large value of $\Omega_R$.
From the data in \cite{todorov2010}, and in a very simplified picture, it is clear that $\tilde{\omega}$, and not $\omega_{12}$, is the electronic resonance to which the electromagnetic cavity seems to couple, resulting in a blueshift of both polariton energies. As a result, at THz frequencies, the coupling ratio $\Omega_R/\tilde{\omega}$ can be smaller than $\Omega_R/\omega_{12}$ by a significant amount, meaning that the depolarization shift is a limiting factor in achieving ultra strong coupling.

We showed recently that because their energy spectrum consists of equidistant levels, quantum wells with a parabolic potential profile \cite{westervelt} allow the observation of strong coupling in the terahertz up to room temperature\cite{geiser}. In addition, it was shown by Kohn~\cite{kohn} that for an optical transition in an ideal parabolic potential, all electron-electron interactions, including the depolarization shift, exactly cancel each other such that $\tilde{\omega} = \omega_{12}$.
The theorem has been confirmed experimentally for parabolic quantum wells ~\cite{brey} by terahertz absorption measurements. As a result, the observable electronic resonance frequency $\tilde{\omega}$ does not increase with increasing doping, making parabolic potentials attractive to study the ultra strong coupling regime. In fact, by expressing the Rabi frequency $\Omega_R$ in terms of $\tilde{\omega}$, the resulting ratio $\Omega_R / \tilde{\omega}$ for any modulation doped parabolic quantum well is
\begin{equation}
\Omega_{R}/\tilde{\omega}=\frac{\sqrt{f_w}}{2},
\label{eq:splitting}
\end{equation}
where $f_w$ is the fraction of the total cavity thickness occupied by the electron gas. A maximum of $\Omega_{R}/\tilde{\omega}=\frac{1}{2}$ is then reached for an electron gas occupying the whole cavity. The parabolic potential created by a strong perpendicular magnetic field on two-dimensional electron gases has also been suggested as a way to reach even larger values of coupling~\cite{Hagenmuller:2010p1619}. 

However, the implications of Kohn's theorem are non trivial in the strong coupling regime, as the position of the polariton eigenstates instead of the peak intersubband absorption energy is measured. In particular, we show here that the simplistic approach that interprets Kohn's theorem as the cancellation of all electron-electron interactions and in which cavity polaritons result from the coupling between an unshifted "bare" transition $ \omega_{12}$ and the cavity is not accurate, as it fails in particular to predict the emergence of the polaritonic gap. In contrast, a better description is one where the depolarization shift is explicitly taken into account, separated from the other electron-electron contributions  that are then lumped into a renormalized transition~\cite{todorov2010}. 

\begin{figure}
\includegraphics[width=0.5\textwidth]{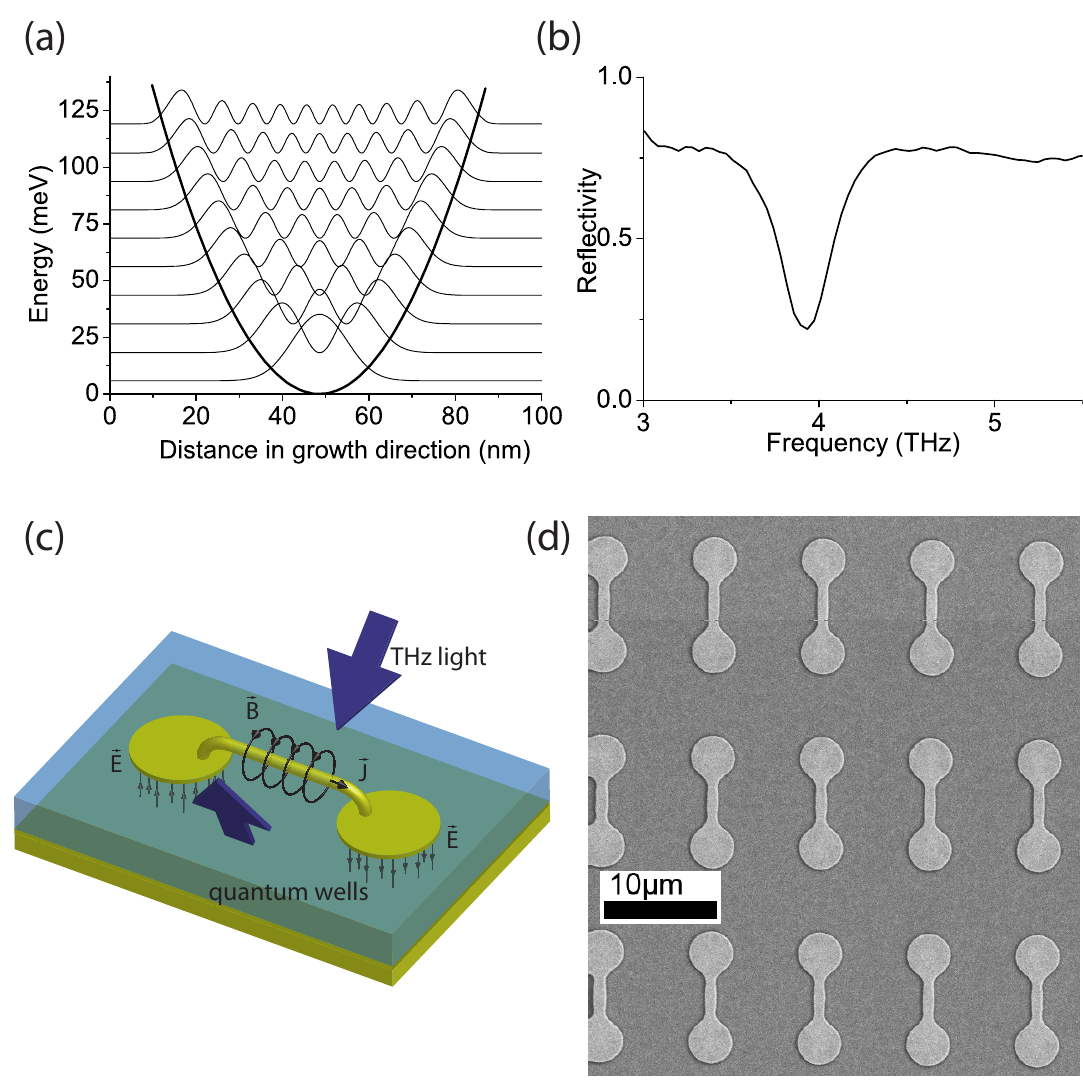}
\caption{\label{fig01}(a) Effective parabolic potential and eigenstates. The potential is digitally alloyed in GaAs / Al$_{0.15}$Ga$_{0.85}$As. The well is surrounded by Al$_{0.3}$Ga$_{0.7}$As barriers, containing a Si-$\delta$-doping. (b) Intersubband absorption in a single pass prism at a temperature of 10K. (c) Schematic of the electronic feedback microcavity, composed of extended circular capacitor elements and a wire - having a finite inductance - connecting them. The cavity can be tuned by changing the electrical size of the components. THz probe light is incident from the top side at a $45^\circ$ angle from the surface normal. (d) Scanning electron microscope picture of a cavity array.}
\end{figure}

A parabolic potential, as shown in fig. \ref{fig01}(a), is approximated by digitally alloying GaAs and Al$_{0.15}$Ga$_{0.85}$As layers.  In contrast to our previous work~\cite{geiser}, and to preserve the translational invariance necessary for the application of Kohn's theorem, we use here a modulation-doped sample where the dopants are set remotely from the active quantum well, reducing greatly the strength of the random impurity potential. The thick line shows the effective potential while the thin lines show moduli squared of the eigenfunctions. Eight parabolic wells with a thickness of $72$nm each are grown by molecular beam epitaxy, separated by Al$_{0.3}$Ga$_{0.7}$As barriers. The total cavity thickness is $t = 0.88 \mu m$ including the contact layers. The structure is silicon $\delta$-doped between the wells to a sheet carrier density of $3.2\times10^{11}cm^{-2}$ per well, as determined in a capacitance-voltage measurement. 
A higher doped sample was also grown, with, for a similar sample thickness ($t = 0.93 \mu m$), 13 quantum wells each doped to a sheet electron density equal to $4.5\times10^{11}cm^{-2}$, as determined by a Shubnikov-de-Haas measurement. From the measured values of electron density and intersubband transition energy, we find $f_w = 0.17$ and $f_w = 0.27$ for the low and higher doped samples, respectively. 

To achieve the small mode volume required for ultra strong coupling, microcavities based on an inductor-capacitor circuit \cite{LC, geiser} were used. They are schematically shown in fig. \ref{fig01}(c).
The quantum wells are sandwiched between the cavities and the gold ground plane. To tune the cavity resonance $\omega_c$, we fabricated multiple fields of resonators with different electrical sizes of the inductance. A top view scanning electron microscope image of a cavity array is shown in fig. \ref{fig01}(d). The electric field of the cavity modes is mainly oriented along the growth direction and therefore couples well to the intersubband transitions, as finite element simulations show.

To fabricate the samples, the epitaxial wafer is Ti/Au-metalized and wafer bonded to a hosting substrate. After substrate removal, resonators are defined by optical lithography, Ti/Au-evaporation and lift-off. Samples of $1$mm$\times 1$mm size, containing several hundred resonators were cleaved and indium soldered to copper heatsinks. Slightly thinner samples of similar but undoped structures as used in \cite{geiser} were used for empty cavity measurements. The cavity resonance frequency dependence on thickness is very weak.

Reflection spectra were taken with a Fourier-transform infrared spectrometer. Light of its glowbar source was focused onto the sample at a $45^\circ$ angle of incidence relative to the normal of the sample surface from the top. Reflected light was refocused onto a helium-cooled Si-bolometer. The light was not polarized and the signal referenced to a gold mirror.
Samples for intersubband absorption measurements were cleaved from the original metalized substrate and polished at a $45^\circ$ angle into single pass prisms with the quantum wells and the metalization on the back. The corresponding spectra were also taken in the above described geometry. All measurements were performed at a temperature of $10$K.

\begin{figure}
\includegraphics[width=0.48\textwidth]{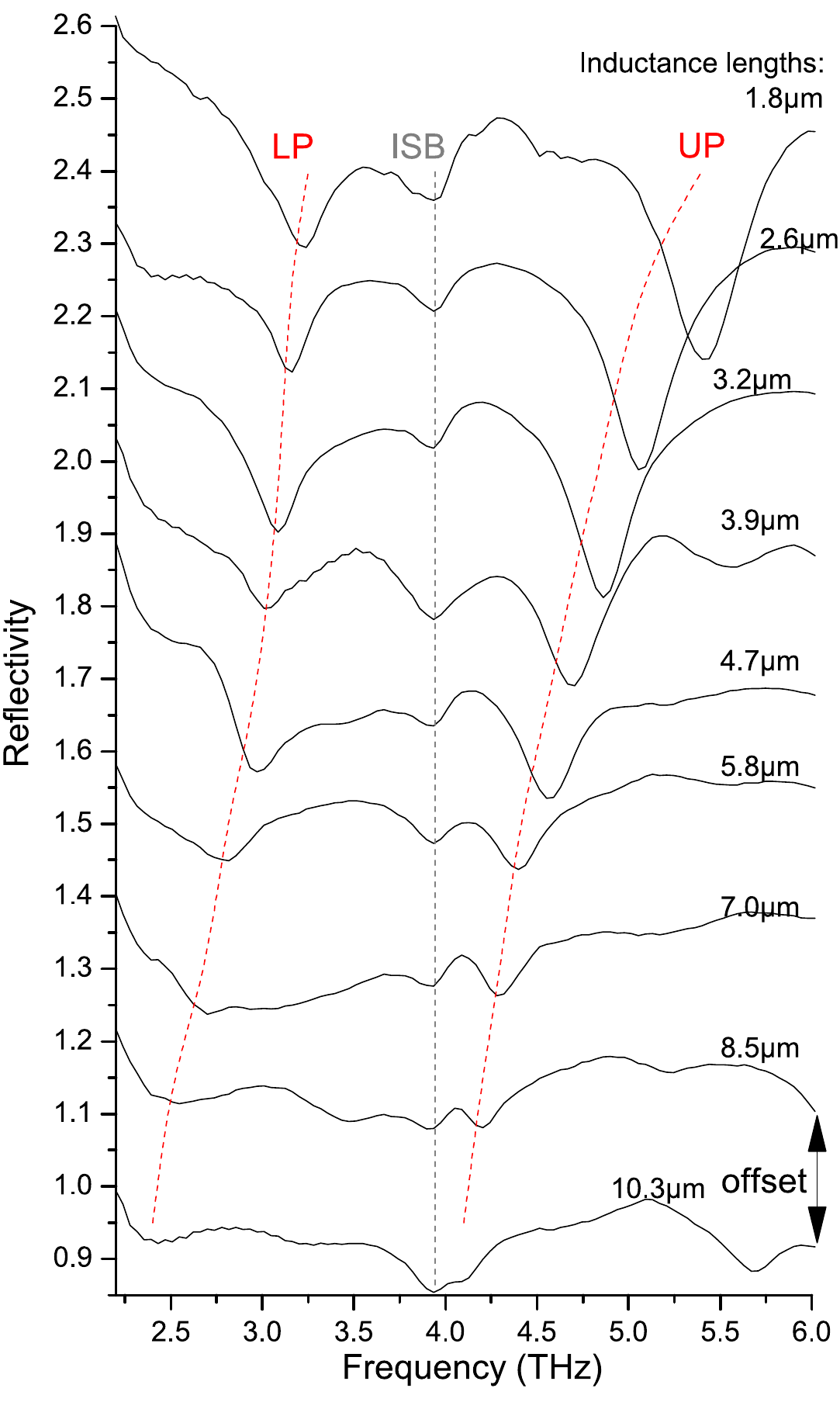}
\caption{\label{fig02} (a) Reflection spectra of samples with different cavity resonance frequencies. The inductor lengths of the cavities are indicated in the graph, while the capacitance is kept constant. The spectra are offset from each other by the indicated amount, with lower spectra having lower cavity resonance frequencies. The three prominent absorption features correspond to the lower polariton (LP), the uncoupled intersubband absorption (ISB) and the upper polariton (UP); the dashed lines are guides to the eye.} 
\end{figure}

Reflection spectra from samples with different cavity resonance frequencies are shown in fig. \ref{fig02} for the lower doped sample, which shows the cleanest absorption data. The lengths of the inductances are indicated in the figure. The capacitance parts had a fixed diameter of $3.3 \mu m$. The spectra show three absorption dips, corresponding to the lower and upper polariton branches (marked LP and UP) and the bare intersubband absorption (marked ISB). In higher spectra (cavity blueshifted from the intersubband transition), the light and matter states are not fully mixed and the line shapes of the polariton resonances are different. As $\omega_c$ is tuned to resonance with the intersubband transition (middle spectra), lineshapes become similar and a strong anticrossing of $2\Omega_R=2\pi \times 1.6$THz is observed. When $\omega_c$ is redshifted from the intersubband transition (lowest spectra), the states decouple again.
The third peak situated between the two polariton branches is found at a constant $\omega = 2 \pi \times 3.9$THz. This absorption feature was also observed in an otherwise identical sample devoid of resonators. We attribute it to intersubband absorption in the bare epitaxial layer between the resonators. The frequency agrees with an intersubband absorption measurement in a single pass prism, shown in fig. \ref{fig01}(b). Similar spectra with persisting resonance center frequencies are obtained at $300$K temperature. They differ in a strongly increased intersubband transition linewidth but are characterized by the same anticrossing frequency $\Omega_R= 2 \pi \times 1.6$THz. Similar results are observed for the heavier doped sample, but with an intersubband absorption peaking at a larger frequency of $\tilde{\omega} = 2 \pi \times 4.5$THz and an even stronger anticrossing frequency $2\Omega_R= 2 \pi \times 2.4$THz.
The blueshift of the intersubband absorption between the measured (4.5THz) and designed (3.8THz) resonant frequency is explained by the fact that the excited states start to "feel" the edge of the parabolic well~\cite{imperfectparabolicwells} for this heavily doped sample, enabling a departure from the prediction of Kohn's theorem. Indeed, a complete self-consistent computation, including the multiple transitions in a mean-field approximation (Hartree term and depolarization shift) predicts a transition energy at 4.55THz, close to the experimentally observed value. Image charge effects caused by the waveguide metal which also violate Kohn's theorem are estimated to be three orders of magnitude less than the polariton linewidth.

The  Rabi frequency values compare well with the ones computed~\cite{ciuti05,todorov2010,geiser}  ($\Omega_R = 2 \pi \times 816$Ghz and $1.17$THz, respectively) assuming the measured doping values as well as an oscillator strength $f = 1$. 

The use of parabolic quantum wells enables us to achieve very large values of the ratio of  $\Omega_{R}/\tilde{\omega} = 0.205$ for the lower doped sample and $\Omega_{R}/\tilde{\omega} = 0.27$ for the heavier doped one. They are larger than the one recently reported for the case of square wells $\Omega_{R}/\tilde{\omega} = 0.2$ \cite{todorov2010}. 

\begin{figure}
\includegraphics[width=0.48\textwidth]{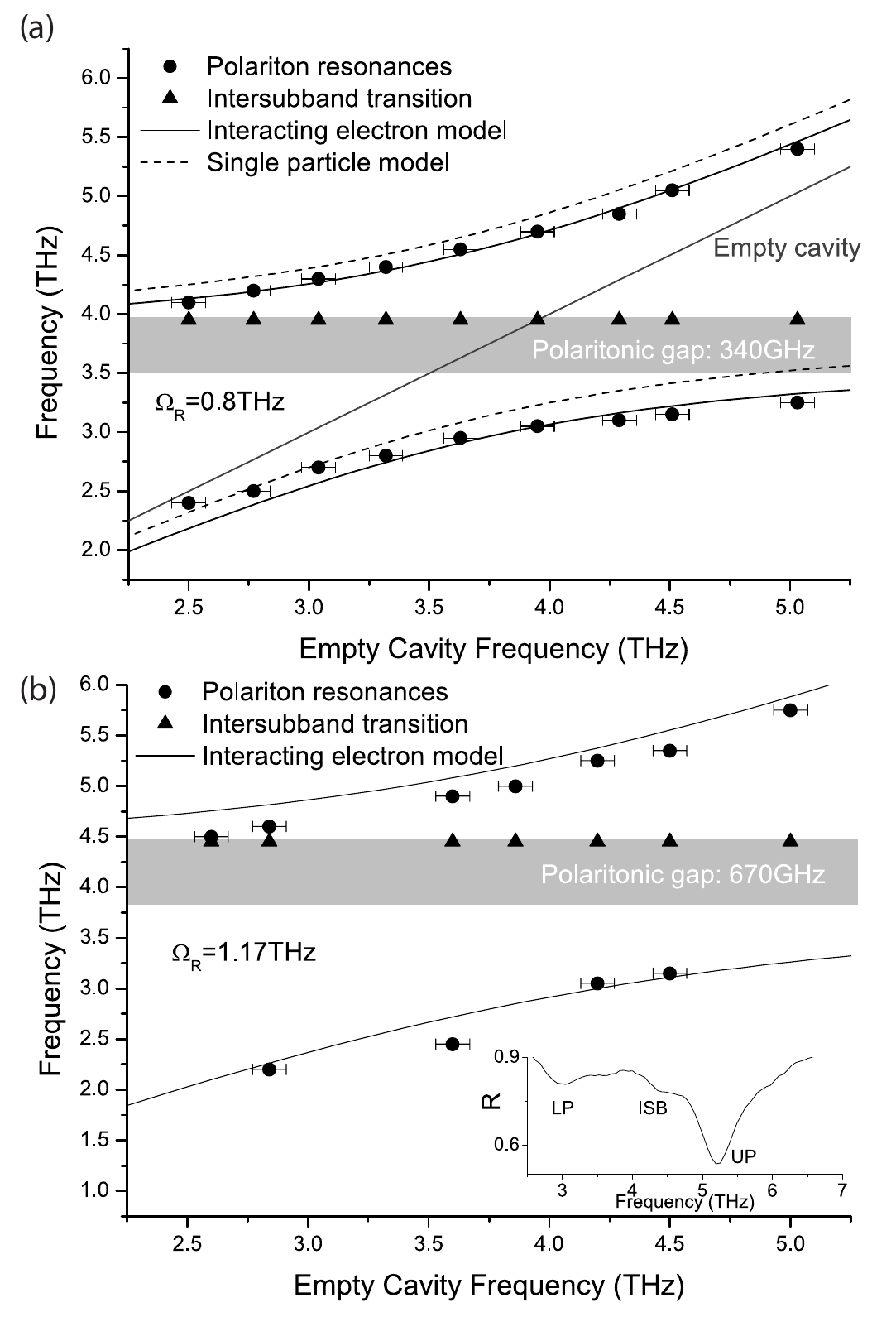}
\caption{\label{fig03} Upper and lower polariton absorption resonances, plotted as a function of the measured empty cavity frequency as dots. The measured uncoupled cavity resonance and intersubband absorption (triangles) are also shown. (a) Lower doped sample. (b) Higher doped sample. Computed points using a Hopfield model (full line) as well as a dielectric model (dashed line, for the lower doped sample only) are shown. The inset in (b) shows a typical absorption spectrum of the higher doped sample.
}
\end{figure}

Fig. \ref{fig03} shows a comparison between our data, for both samples, with two different models, both considering a single intersubband transition strongly coupled to an electromagnetic cavity mode and taking into account the resonant and antiresonant electron-photon interaction terms.  Dashed lines are computed using the approach reported by Ciuti et al.\cite{ciuti05} where the intersubband plasmon is coupled to the cavity. In solid line, we show the predictions of the model reported by Todorov et al.~\cite{todorov2010} in the dipolar gauge, where the depolarization shift is included as a quadratic polarization term in the electron-photon interaction Hamiltonian. Such a term plays the role of an electron-electron interaction term leading to bound state energy renormalization, and it was shown to be essential to correctly reproduce the experimental results~\cite{todorov2010} including the prediction of the existence of a polariton gap. The Coulomb gauge and the dipolar gauge are connected by a unitary transformation. In this model, the intersubband plasmon energy $\tilde{\omega}$ is given by  $\tilde{\omega}^2 = \left( \omega_{12}^{ee}\right)^2 + \omega_P^2$, with $\omega_{12}^{ee}$ being the intersubband frequency that includes all electron-electron interactions (in particular Hartree and correlation) except the depolarization shift. In the square well case of \cite{todorov2010}, this leads to a strongly renormalized $\tilde{\omega}$. In the present case of a parabolic potential, the frequency $\omega_{12}$ is heavily reduced to almost zero as seen in a selfconsistent Schr\"odinger-Poisson calculation which shows a strongly flattened potential due to the electron charge density. Furthermore, $\omega_P$ already accounts for the measured intersubband plasmon frequency, justifying $\omega_{12}^{ee} = 0$. This can also be enforced as a consequence of Kohn's theorem.
Both models reproduce the data fairly well. However, the agreement with the data of the model presented by Todorov~\cite{todorov2010} is better and it reproduces the polaritonic gap of size
\begin{equation}
\omega_{gap} \approx f_w \frac{\omega_P^2}{2 \tilde{\omega}_{12}} = f_w \omega_P /2 .
\end{equation} 
equal to 340GHz for the lower doped sample and 670Ghz for the higher one. 

In conclusion, in the strong and ultra strong coupling regime a more complete picture of the meaning of Kohn's theorem arises: electron-electron interaction terms cancel each other's effects on the resonance frequencies, but a more accurate modeling shows that the individual terms maintain physical meaning, as the plasma excitation still opens a polaritonic gap.

 In addition to these insights, our findings also have important practical implications for the realization of ultra-strong coupling as the compensation of a depolarization shift indicates parabolic quantum wells as the systems in which the highest $\Omega_{R}/\tilde{\omega}$ ratios can be observed.

We would like to thank Christian Reichl and Werner Wegscheider for supporting measurements.
This work was supported by the Swiss National Science Foundation through contract no $200020\_129823/1$.

\end{document}